\documentclass[preprint,amsmath,amssymb,superscriptaddress]{revtex4}
\usepackage{dcolumn}
\usepackage{bm}
\usepackage{graphicx}

\begin{document}

\title{The effect of a market factor on information flow between stocks using minimal spanning tree}

\author{Cheoljun Eom}
\affiliation{Division of Business Administration, Pusan National University, Busan 609-735, Republic of Korea}
\email{shunter@pusan.ac.kr}

\author{Okyu Kwon}
\affiliation{Northwestern Institute on Complex Systems, Northwestern University, Evanston,, IL 60208, USA}

\author{Woo-Sung Jung}
\affiliation{Department of Physics and Basic Science Research Institute, Pohang University of Science and Technology, Pohang 790-784, Republic of Korea}
\affiliation{Asia Pacific Center for Theoretical Physics, Pohang 790-784, Republic of Korea}
\affiliation{Center for Polymer Studies and Department of Physics, Boston University, Boston, MA 02215, USA}

\author{Seunghwan Kim}
\affiliation{Department of Physics and Basic Science Research Institute, Pohang University of Science and Technology, Pohang 790-784, Republic of Korea}
\affiliation{Asia Pacific Center for Theoretical Physics, Pohang 790-784, Republic of Korea}

\date{\today}

\begin{abstract}
We empirically investigated the effects of market factors on the information flow created from $N(N-1)/2$ linkage relationships among stocks. We also examined the possibility of employing the minimal spanning tree (MST) method, which is capable of reducing the number of links to $N-1$. We determined that market factors carry important information value regarding information flow among stocks. Moreover, the information flow among stocks evidenced time-varying properties according to the changes in market status. In particular, we noted that the information flow increased dramatically during periods of market crises. Finally, we confirmed, via the MST method, that the information flow among stocks could be assessed effectively with the reduced linkage relationships among all links between stocks from the perspective of the overall market.
\end{abstract}

\maketitle

\section{Introduction}
The financial market is known to create close linkage relationships among assets and markets. This can be observed in cases in which a financial crisis occurring initially in one country exerts influence on other countries, and the bankruptcy of one company can similarly affect not only related stocks, but simultaneously, other stocks that are unrelated to the particular company within the domestic stock market. Accordingly, examining the information flow arising from interactions among assets and markets in a financial market is a crucial research topic for our understanding of the pricing mechanism. Currently, many researchers in the field of finance, as well as those in the discipline of econophysics, are expending significant effort to characterize information flow in financial markets.

Unlike previous studies, our objective in examining information flow among stocks was not to verify the lead-lag relationship between selected stocks. Our goal is to discover empirically the influential factors that can affect the existence of significant information flow among stocks from the perspective of the overall market. The stocks traded on the stock market initiate interactions on the basis of external forces (inflow of information), and interactions among stocks induce exchanges and transfers of information, and then establish the pricing mechanism. Therefore, assessing the information flow regarding the interactions among stocks in a stock market is helpful in understanding the pricing mechanism. However, there have not been many studies thus far conducted that have dealt with the factors that affect information flow among stocks from the overall market perspective.

In a recent study by Eom \emph{et al.}, the Granger causality model was utilized to propose time scales of return measurement and differences in efficiencies among stocks with the influential factors that affect the existence of significant information flow between stocks, statistically [2]. On the basis of the time dependency properties of financial time series data [3], Eom \emph{et al.} empirically demonstrated that extending the period for converting price into return reduces the amount of significant information flow among stocks. Furthermore, they demonstrated that different efficiencies among stocks influence the direction of information flow on the basis of the efficient market hypothesis [4], which claims that there are discrepancies in the response speeds of stocks in a stock market when information is introduced to the market. They utilized the approximate entropy method as a measurement for the quantification of the degree of efficiency [5-7]. On the basis of their research approach, we intend to empirically investigate other influential factors that may affect the existence and magnitude of significant information flow among stocks.

In this study, we considered market factors as potential elements that may affect the information flow among stocks. In previous studies, it was known that market factors carry critical information values regarding changes in stock prices. King suggested market, industrial and company factors as common factors that can commonly explain price changes in the stocks traded on the market, and proved that the market factors are the most influential of these [8]. His study results affected the capital asset pricing model [9-11] and the arbitrage pricing model [12], both of which are representative pricing models in the finance field. It is generally recognized in the field of finance and econophysics that the largest eigenvalue elicited from principal component analysis and the random matrix theory method has market factor properties [13-17]. It has also been demonstrated that the properties of the largest eigenvalue have identical market factor attributes, regardless of the number and types of stocks included in the sample [18]. Additionally, it has been demonstrated in previous studies that the degree of commonality of the market factors varies in accordance with market status, and particularly increase during times of market crises. King claimed that the ratio of stock price changes that could be explained by market factors increased during the American Great Depression in the 1920s [8]. Onnela \emph{et al.} determined that the sizes of correlation matrices among stocks increased precipitously in the period around Black Monday in 1987 in the US [19-20]. Eom \emph{et al.} provided empirical evidence showing that the size of the largest eigenvalue increases during market crises [21]. On the basis of the results of the aforementioned studies, we evaluated the effects of market factors on the information flow among stocks in terms of two perspectives-namely, information value and changes in market status.

Although every linkage relationship with interactions among stocks should be utilized for investigations in order to assess information flow among stocks from the perspective of the overall market, such testing processes require tedious calculations. There are $N(N-1)/2$ links among all of the stocks traded on the market. Using these links, we can determine whether the information flow among stocks is a mutual exchange of information, a one-way information transfer, or an absence of information flow. Because we utilized 104 and 310 stocks in the Korean and US stock markets, respectively, there are 5,356 (=$(104 \times 103)/2$) and 47,895 links among stocks that must be evaluated from the viewpoint of the overall market. In an effort to conduct a more effective examination of information flow among stocks from the overall market perspective, we required a research approach that could effectively reduce the number of links among stocks. For this purpose, we proposed the minimal spanning tree (MST) method. One of the concepts that have been recognized as significant achievements among disciplinary research efforts in the field of finance and econophysics is the stock network originally proposed by Mantegna [22]. In order to gain a better understanding of the pricing mechanism, this method selects only the stocks with the closest interactions among all stocks, and generates a visual presentation of the linkage relationships among selected interactions between stocks. The correlation matrix is utilized as a means for the quantitative measurement of the interactions among stocks. From the $N(N-1)/2$ correlation matrix elements of all stocks, $N-1$ with significant interactions are elicited via the MST method [23-25]. Accordingly, we intend to verify whether the links reduced from all of the linkage relationships among the stocks using the MST method are capable of effectively characterizing the information flow among stocks from the perspective of the overall market.

As mentioned above, our research objective is to empirically assess the effects of market factors on the existence of information flow among stocks from the perspective of the overall market. Moreover, with the objective of proposing an effective research methodology of the testing process, we intend to determine whether information flow can be observed only with the links reduced via the MST method from the perspective of the overall market. The observed results can be summarized as follows. We determined that market factors have important information value with regard to the information flow among stocks. When the market factors' properties included in stock returns were eliminated, the amount of significant information flow among stocks was reduced substantially. Moreover, the quantity of significant information flow among stocks varied in accordance with changes in the market status. In particular, the quantity of significant information flow among stocks increased dramatically during market crises. Additionally, we were able to determine that using only $N-1$ links via the MST method, one could generate similar information results as were observed using all $N(N-1)/2$ links among stocks. In other words, we confirmed that the MST method is an effective means of observation of information flow among stocks from an overall market perspective.

This paper is constructed as follows. After the introduction, the subsequent chapter explains the data and methods used to empirically investigate the established study objectives. Chapter III presents the results observed in accordance with the empirical design, and the final chapter summarizes our observed results and provides some of the implications of those results.

\section{Data and methods}
\subsection{Data}
We selected stocks from Korean and US stock markets with consecutive daily price data during the 18-year period from January 1990 to December 2007, respectively. There were 104 stocks from the KOSPI 200 index of Korean stock market and 310 from the S\&P 500 market index of US stock market. Additionally, in order to assess the changes in information flow among stocks around market crises, we compiled the data of 109 stocks with consecutive daily price data during the 12-year period from January 1992 to December 2003, which includes the Asian Foreign Exchange Crisis of 1997.

As for the comparative reference for the evaluation of the effects of market factors on information flow among stocks, we utilized the results of information flow among stocks according to changes in time scales for the measurement of the returns noted in the previous study [2]. According to the results of the previous study, since the results regarding information flows among stocks are directly affected by changes in the time scales of return measurement, we must consider the results from the changes in the time scale of return in order to clearly assess whether the observed results were caused by market factors. The time scales of return are the time intervals utilized to convert the price data into return data. In other words, $k=1,2,\dots,5$ is used in $R_k(t)=\ln P(t)-\ln P(t-k)$. $R_1(t)$ is a case in which $k=1$ is utilized as the time interval, corresponding to the daily returns. $P_5(t)$ uses $k=5$ as the time interval, which corresponds to a weekly return. As there are insufficient data for use in the analysis of $R_k(t)$ for $k\ge6$, we established the maximum time interval at $k=5$.

\subsection{The Granger causality model}
This section explains the Granger causality model, which is capable of assessing significant information flow among stocks from a statistical viewpoint. This model determines the direction of information flow among time series data on the basis of statistical significance, in order to determine whether fluctuation in specific time series data can explain variation in another set of time series data. Using the Granger causality model and all $N(N-1)/2$ links with interactions among stocks, we will assess the significant information flow among stocks from a statistical perspective. As we used 104 and 310 stocks in Korean and US stock markets, respectively, there were 5,356 (=$(104 \times 103)/2$) and 47,895 links among stocks that required examination.

\begin{eqnarray}
H_0:&&R_B \textrm{ does not Granger-cause } R_A  \nonumber \\
&&R_B \Rightarrow R_A(k,L):~ R_{A,k(t)}=C_1+\sum^{L}_{\tau=1}\alpha_{1,\tau}R_{A,k(t-\tau)}+\sum^{L}_{\tau=1}\beta_{1,\tau}R_{B,k(t-\tau)} \\
H_0:&&R_A \textrm{ does not Granger-cause } R_B  \nonumber \\
&&R_A \Rightarrow R_B(k,L):~ R_{B,k(t)}=C_2+\sum^{L}_{\tau=1}\alpha_{2,\tau}R_{B,k(t-\tau)}+\sum^{L}_{\tau=1}\beta_{2,\tau}R_{A,k(t-\tau)} \\
\textrm{where}&&k=1,2,\dots,5 \textrm{ and } L=1,2,\dots,5 \nonumber
\end{eqnarray}

$R_{A,k(t)}$ and $R_{B,k(t)}$ are the returns of stock $A$ and $B$ converted with the time interval $k$, respectively, and $\alpha$ and $\beta$ are regression coefficients, and $C$ is a constant. In order to explain the fluctuation in returns calculated at $t$ with the time interval $k$, $R_B \Rightarrow R_A(k,L)$ in Eq. 1 and $R_A \Rightarrow R_B(k,L)$ in Eq. 2 utilize the second term on the right, its own past return data of a dependent variable from $t-1$ to $t-L$, and the third term, the past return data of another stock from $t-1$ to $t-L$, as independent variables. As in the study of Eom \emph{et al.}, we employed the past return data from a minimum of $L=1$ to a maximum of $L=5$. Because information flow among stocks was tested using the past data up to a maximum of $L=5$ for the return data of the maximum time interval $k=5$, Eq. 1 and 2 were repeatedly tested for a total of 25 [$(k=5)\times(L=5)$] cases.

From all $N(N-1)/2$ linkage relationships, we categorized the significant information flow among stocks into two types of mutual exchange and one-way direction of information flow. Naturally, links not associated with two types of information flow are cases of no exchange of information. Granger F-statistics (Wald statistics) at a significance level of 5\% was utilized to statistically determine the amount of significant information flow. In other words, if the Granger F-statistics are statistically significant, $R_A \Rightarrow R_B(k,L)$ represents information transfer from $R_A$ to $R_B$, and significant $R_B \Rightarrow R_A(k,L)$ from $R_B$ to $R_A$. Accordingly, if $R_A \Rightarrow R_B(k,L)$ and $R_B \Rightarrow R_A(k,L)$ are both significant, the type of information flow between stocks $A$ and $B$ is a mutual exchange of information flow, $A\rightleftharpoons B$. On the other hand, if either $R_A \Rightarrow R_B(k,L)$ or $R_B \Rightarrow R_A(k,L)$ is significant, it constitutes a one-way direction of information flow, $A\rightarrow B$ or $A\leftarrow B$. Meanwhile, we shall assess the changes in the amount of significant mutual exchange and one-way direction of information flow. Among all $N(N-1)/2$ links, we utilized the ratio of  $FR(k,L:A\rightleftharpoons B)^T = \frac{f_q(k,L:A\rightleftharpoons B)^T}{N(N-1)/2}$ incorporated with the amount of significant mutual exchange of information flow, $f_q(k,L:A\rightleftharpoons B)^T$, and $FR(k,L:A\rightarrow B)^T$ ratio with the number designating the significant one-way direction of information flow, $f_q(k,L:A\rightarrow B)^T$, in which superscript T denote a case in which all $N(N-1)/2$ links are utilized.

\subsection{Minimal spanning tree method}
This section explains the MST method for reducing the number of links among stocks in order to conduct a more effective examination of the information flow among stocks from the perspective of the overall market [23]. This method is similar to the single linkage method of cluster analysis, which is a multivariate statistical analysis [26]; the MST method utilized the correlation matrix, $\rho_{A,B,k}$ , which contains $N(N-1)/2$ elements, as a measurement of interactions among stocks.

\begin{equation}
\rho_{A,B,k}=\frac{\left< R_{A,k(t)}R_{B,k(t)} \right> -\left< R_{A,k(t)} \right> \left< R_{B,k(t)} \right> }{\sqrt{ \left( \left< R_{A,k(t)} \right> - \left< R_{A,k(t)} \right>^2 \right) \left( \left< R_{B,k(t)} \right> - \left< R_{B,k(t)} \right>^2 \right) }} ~~\left(-1 \le \rho \le 1 \right)
\end{equation}

Here, $\left< \cdots \right>$ indicates the average of return at the time interval $k$. In order to select the links with high correlation among stocks in topological space, $\rho_{A,B,k}$ are used to compute the distance matrix $d_{A,B,k}$ among stocks:

\begin{equation}
d_{A,B,k}=\sqrt{ 2(1-\rho_{A,B,k})}
\end{equation}

On the basis of $d_{A,B,k}$, the closest $N-1$ links are selected from $N(N-1)/2$ links using the Kruskal algorithm [27], which is a graph without a cycle connecting all nodes with links. Here, a smaller (greater) $d_{A,B,k}$ ($\rho_{A,B,k}$) value is evaluated in order to represent a closer relationship.

Using $N-1$ links reduced via the MST method, we evaluated the amount of significant mutual exchange of information flow, $f_q(k,L:A\rightleftharpoons B)^{MST}$, and significant one-way direction of information flow,  $f_q(k,L:A\rightarrow B)^{MST}$, and calculated the ratios $FR(k,L:A\rightleftharpoons B)^{MST} = \frac{f_q(k,L:A\rightleftharpoons B)^{MST}}{(N-1)}$ and $FR(k,L:A\rightarrow B)^{MST}$, respectively, in which superscript MST denotes the case using $N-1$ links. In addition, we determined whether identical results could be acquired between changes in $FR(.)^T$ and $FR(.)^{MST}$. On the basis of our comparison between results, we were able to determine whether the MST method is an alternative means for effectively assessing the information flow among stocks from the overall market perspective.

\section{Result}
\subsection{The influential factors: Time dependency properties}
In this section, we present the results of the effects on the information flow among stocks according to changes in the time scale of return as confirmed in previous studies, prior to examining the effects of market factors on the information flow among stocks. Because it has been verified that the information flow among stocks is affected directly by the changes of time scale in return, we must consider these results in order to clearly confirm that the observed result from the objective in this study is induced by market factors. The observed results are displayed in Fig. 1.

Fig. 1(a) and (b) show the average ratio of significant information flow among stocks, $\overline{FR(k)^T}=\frac{1}{5}\sum^5_{L=1}FR(k,L:.)^T$, $k=1,2,\dots,5$, observed using all $N(N-1)/2$ links among stocks. Fig. 1(c) and (d) show the average ratio of the significant information flow among stocks, $\overline{FR(k)^{MST}}$ observed using only $N-1$ links reduced via the MST method. Significant mutual exchange of information flow is indicated by red circles for Korean stock market and red triangles for US stock market, and significant one-way direction of information flow is designated by blue squares (Korea) and blue pentagrams (US). Meanwhile, in an effort to confirm that the observed results are characteristics observed only from financial time series data, the results obtained from an identical testing process using shuffled time series and white noise data are provided in Fig. 1 (b) and (d).

According to the results observed herein, time scale of return is apparently a influential factor that affects the results of significant information flow among stocks, regardless of the number of links employed [ $N(N-1)/2$ and $N-1$]. In other words, as the time interval for converting price data into return data is extended from $t=1$ to $t=5$, the average ratios of significant mutual exchange and one-way direction of information flow all decline [Fig. 1 (a) and (c)]. On the other hand, significant changes cannot be observed according to the results from the shuffled time series data and the white noise data [Fig. 1 (b) and (d)]. One point of interest is that the average ratios of significant information flows observed from $N-1$ links via the MST method have higher values than those observed from all $N(N-1)/2$ links--that is to say, $\overline{FR(k)^{MST}}>\overline{FR(k)^T}$ . Such a result can be inferred from the fact that links with a strong correlation between stocks evidence more significant information flow than links with a weak correlation between stocks. In order to empirically confirm this inference, we divided the $N(N-1)/2$ correlations between stocks into links with the upper 20\% correlation and those with the lower 20\% correlation in terms of the size of correlation value, and then compared the numbers of significant information flow for the two groups. The observed results are provided in Fig. 2. Figs. 2(a) and (b) provide the results using data from Korean and US stock markets, respectively. In the group with the upper 20\% correlation, significant mutual exchange and one-way direction of information flow are indicated with red circles and blue squares, respectively, and the group with the lower 20\% correlation is designated by red triangles and blue pentagrams. According to the results, the average ratio of significant information flow for the group in the upper 20\% correlation was higher than that of the lower 20\% group. These results indicate that since the MST method reduces   links into   links based on the degree of closeness of the relationship in terms of the size of correlation between stocks, the ratio of significant information flow from the links reduced via the MST method is greater than that from all links.

\subsection{The influential factors: Market factor}
This section presents the results observed for the effects of market factors on the existence of significant information flow among stocks, statistically. As mentioned in previous studies [8, 13-14, 16-18], if a market factor has information value concerning stock price changes, that market factor can be an influential factor that can exert critical effects on the information flow among stocks. Two methods can be employed to verify the effects of market factors. One of these methods is to introduce market factors as new independent variables to the model, and the other is to remove the properties of the market factors from the original stock return data. The observed results shown in Fig. 1 should contain data concerning market factors. Accordingly, we shall generate data by removing the properties of market factors from the original stock return data and compare the results observed with the testing process identical to the one utilized in Fig. 1.

Time series data with market factor properties generally use the actual market index. However, as the actual market index incorporates every stock traded in the stock market, whereas our data only use stocks in a sample group that satisfies the given conditions, we utilized the time series data of the largest eigenvalue generated via the random matrix theory (RMT) method [15-18]. It is widely known not only in the field of finance but also in the field of econophysics that the time series data reflecting the properties of the largest eigenvalue evidence the properties of a market factor. Via the RMT method, the time series data reflecting the properties of the largest eigenvalue $\lambda_{i=1}$ were calculated on the basis of $R_{M,t}=\sum^N_{j=1}V_{(i=1),j}R_{j,t}$, $t=1,2,\dots,T$ where $V_{(i=1),j}$ is the eigenvector reflecting the properties of $\lambda_{i=1}$, and $R_{j,t}$ is the actual return on stock $j$. The generated $R_{M,t}$ is converted into return data $R_{M,k(t)}$ according to time interval $k$ as established in Section 2 and employed in the subsequent testing process.

Now we shall explain the process of removing the market factor properties from the original stock return data using $R_{M,k(t)}$. The first step involves estimating the regression coefficients $\widehat{\alpha_{j,k}}$ and $\widehat{\beta_{j,k}}$ from the regression equation $R_{j,k(t)}=\widehat{\alpha_{j,k}}+\widehat{\beta_{j,k}}R_{M,k(t)}+\epsilon_{j,k(t)}$ that utilizes $R_{M,k(t)}$ as a independent variable, in order to explain the fluctuation in the actual stock return $R_{j,k(t)}$ with the time interval $k$. Here, $\epsilon_{j,k(t)}$ is an error term. The second step is the combination of the estimated $\widehat{\alpha_{j,k}}$ and $\widehat{\beta_{j,k}}$ with $R_{M,k(t)}$ to obtain $\widehat{R_{j,k(t)}}=\widehat{\alpha_{j,k}}+\widehat{\beta_{j,k}}R_{M,k(t)}+\epsilon_{j,k(t)}$, from which the estimate stock return $\widehat{R_{j,k(t)}}$ is calculated. $\widehat{R_{j,k(t)}}$ is a time series that sufficiently reflects only the market factor properties. The third step is subtracting $\widehat{R_{j,k(t)}}$  from actual stock return $R_{j,k(t)}$, expressed as $R^*_{j,k(t)}=R_{j,k(t)}-\widehat{R_{j,k(t)}}$, where $R^*_{j,k(t)}$ is the return data with the properties of market factor removed from the original stock return. The three-step process is repeated for return time scales $k=1,2,\dots,5$. $R^*_{j,k(t)}$ is then utilized in the testing process identical to the one employed in Fig. 1.

Fig. 3 displays the results of information flow among stocks observed using the return data without the properties of market factors based on the methods explained above. Figs. 3(a) and (b) are results using $N(N-1)/2$ links and $N-1$ links via the MST method, respectively. The figure shows the average ratios of significant information flows using the returns data without market factor properties, $\overline{MFR(.)^T}$ and $\overline{MFR(.)^{MST}}$. The average ratios of mutual exchange of information flow are indicated with red circles (Korean stock market data) and red triangles (US stock market data), and those of one-way direction of information flow are designated by blue squares (Korea) and blue pentagrams (US).

On the basis of the observation results, we were able to confirm empirically that the market factor properties have important information value for the information flow among stocks. In other words, the average ratio of significant information flow among stocks observed from the time series data without the market factor properties is substantially lower than the average value observed from the original stock return data. Table 1 shows the differences in the average ratios $\overline{DFR(.)} = \overline{FR(.)}-\overline{MFR(.)}$ observed in Fig. 1 and Fig. 3 according to $N(N-1)/2$ and $N-1$ links, as well as the time scale $k$. The difference, $\overline{DFR(.)}$ in the average ratio of significant mutual exchange of information flow for the time scale  $k=1$ from $N(N-1)/2$ links ($N-1$ links) was 19.69\% (25.44\%) for Korea and 4.95\% (4.66\%) for US. As for significant one-way direction of information flow, the differences were 22.80\% (4.85\%) and 16.43\% (9.77\%) for Korea and US, respectively. Although the magnitudes of reduction at other $k=2,\dots,5$ were not as large as in the case of $k=1$, the average ratios from the returns without the market factor properties were lower than the results obtained from original returns for other time scales of return. In particular, the extent of reduction of the average ratio for the mutual exchange of information flow was particularly sizeable. Additionally, the reduction observed from the Korean stock market data was greater than that of the US stock market data. This is believed to be attributable to the fact that the degree of commonality for market factors was higher among the attributes of the stocks traded in the Korean stock market than those in the US stock market.

\subsection{The influential factors: the changes of market status}
This section provides the results obtained by examining the effects on the information flow among stocks according to changes of market status. Previous studies have asserted that the effects of changing market status can be clearly assessed during market crises [8, 19-21]. Accordingly, we utilized a period of 144 months around the Asian Foreign Exchange Crisis of 1997, from January 1992 to December 2003. The data consisted of 109 stocks in Korean stock market's KOSPI 200 index. The method of the time scales of return for the selected data was identical to the one elucidated in Section 2. In order to determine the variation in the information flow among stocks according to the changes of market status in terms of time series, we utilized the rolling sample method. The testing period for evaluating significant information flow among stocks was 48 months and the moving period was 1 month. Accordingly, the information flow among stocks was observed statistically using the data from the initial 48 months, and then the data from the first month were removed, and the data from the 49th month was added for observation using the data from the second set of 48 months. The testing process was repeated 97 (=$144 - 48 + 1$) times. Fig. 4 displays the results observed from examining the time series variation of the information flow among stocks. Figs. 4(a) and (b) are the results observed using $N(N-1)/2$ links and $N-1$ links via the MST method, respectively. The figure shows the average ratio $\overline{FRs}$ of significant mutual exchange and one-way direction of information flow for the two time scales of return at $k=1$ and $k=5$. The mutual exchange and one-way direction of information flow for $k=1$ are designated by red circles and blue squares, respectively, and with cyan triangles and magenta pentagrams, respectively, for $k=5$. The Asian Foreign Exchange Crisis of December 1997 is marked with an `x'.

According to the observed results, we determined that the results of the information flow among stocks according to market status changes have time-varying properties. In other words, the number of significant mutual exchange and one-way direction of information flow among stocks are sensitive to changes in the market status. Accordingly, we conducted the Jarque-Bera (JB) statistical test [28] for a robust observation of the time-varying properties in the results of the information flow among stocks. The test evaluates whether the data departs from normality, and is also utilized as a measurement to determine the time-varying properties of data in the field of finance. The average ratio of mutual exchange of information flow from $N(N-1)/2$ for $k=1$ yielded JB=15.37 and a one-way direction of information flow JB=9.60, indicating that both were significant at a significance level of 1\%. The results from $N-1$ links via the MST method did not differ significantly from the results confirmed by $N(N-1)/2$ links. The average ratio of mutual exchange of information flow yielded JB=6.47 and a one-way direction of information flow of JB=7.20, thereby indicating that both were significant at a significance level of 5\%. Furthermore, at $k=5$, the JB test statistics confirmed by mutual exchange and one-way direction of information flow were significant at a significance level of 1\%, regardless of the number of links. Consequently, the observed results function as robust empirical evidence that the average ratio of information flow evidences time-varying properties according to changes in the market status. We also determined that in Fig. 4, the average ratios of mutual exchange and one-way direction of information flow dramatically increased at the time of the Asian Foreign Exchange Crisis (marked with `x'). On the basis of the findings of previous research and this study, such a finding allows us to infer that a market crisis increases the degree of commonality of the market factor included in the data of stocks traded on the market, increases the interaction among stocks, which enhances the generation of information flow among stocks.

\section{Conclusion}
We empirically investigated the effects of market factors on the existence of significant information flow among stocks from the perspective of the overall market in terms of information value and changes in market status. With the objective of developing a research methodology for the effective observation of information flow among stocks, we attempted to determine whether information flow could be evaluated with links reduced via the MST method. We utilized stock data traded on the Korean and US stock markets in order to robustly confirm the established research objective. The observed results can be summarized as follows.

First, we verified that variation in the time scales of return influence the results of significant information flow among stocks, statistically. Extending the time interval utilized for the conversion of price data into return data reduced the ratios of significant mutual exchange and one-way direction of information flow. Furthermore, the ratio of significant information flows observed from   links via the MST method had a greater value than the result observed from all   links. It was confirmed that this result was attributable to the fact that links with a high correlation among stocks have more significant information flow than links with a low correlation.

Second, we determined that the properties of market factor have important information values for information flow among stocks. The average ratio of significant information flow among stocks observed from the time series data without market factor properties had apparently lower values than the average ratio observed from the original stock return data. In particular, the extent of reduction of the average ratio for significant mutual exchange of information flow was particularly large. Additionally, the reduction observed from the Korean stock market data was greater than that of the US stock market data. This is believed to be attributable to the fact that the degree of commonality for market factors is higher in the Korean stock market than in the US stock market.

Third, we discovered that the amount of significant information flow among stocks according to the changes of market status evidenced time-varying properties. Specifically, we noted that the average ratio of significant mutual exchange and one-way direction of information flow dramatically increased during period of market crisis. These findings permit us to infer that a market crisis increases the degree of commonality of the market factors included in the stocks data, and raises the interaction among stocks, which consequently increases the frequency of generation of information flow among stocks.

From this study, we were able to empirically confirm that market factors have important information values with regard to the information flow among stocks ,and the observed results have time-varying properties according to changes in the market status. Moreover, we learned that the links reduced by the MST method are capable of effectively observing the information flow among stocks from the perspective of the overall market. Such findings confirm the roles of market factors, which have important implications for the information flow among stocks, thus helping us to better understand the pricing mechanism. Additionally, our proposal of rationally reducing the number of links among all stocks in order to confirm the information flow among stocks in a more effective manner constitutes another significant achievement. Although we focused in this study on the orientation of information flow among stocks from a statistics perspective, future studies will need to expand the scope of this research to determine the amount and type of information exchanged and transferred under conditions of significant information flow.

\begin{acknowledgements}
This work was supported by the Korea Science and Engineering Foundation (KOSEF) grant funded by the Korea government (MEST) (No. R01-2008-000-21065-0), and for two years by Pusan National University Research Grant.
\end{acknowledgements}

\newpage
\clearpage

\begin{figure}
\includegraphics[width=1.0\textwidth]{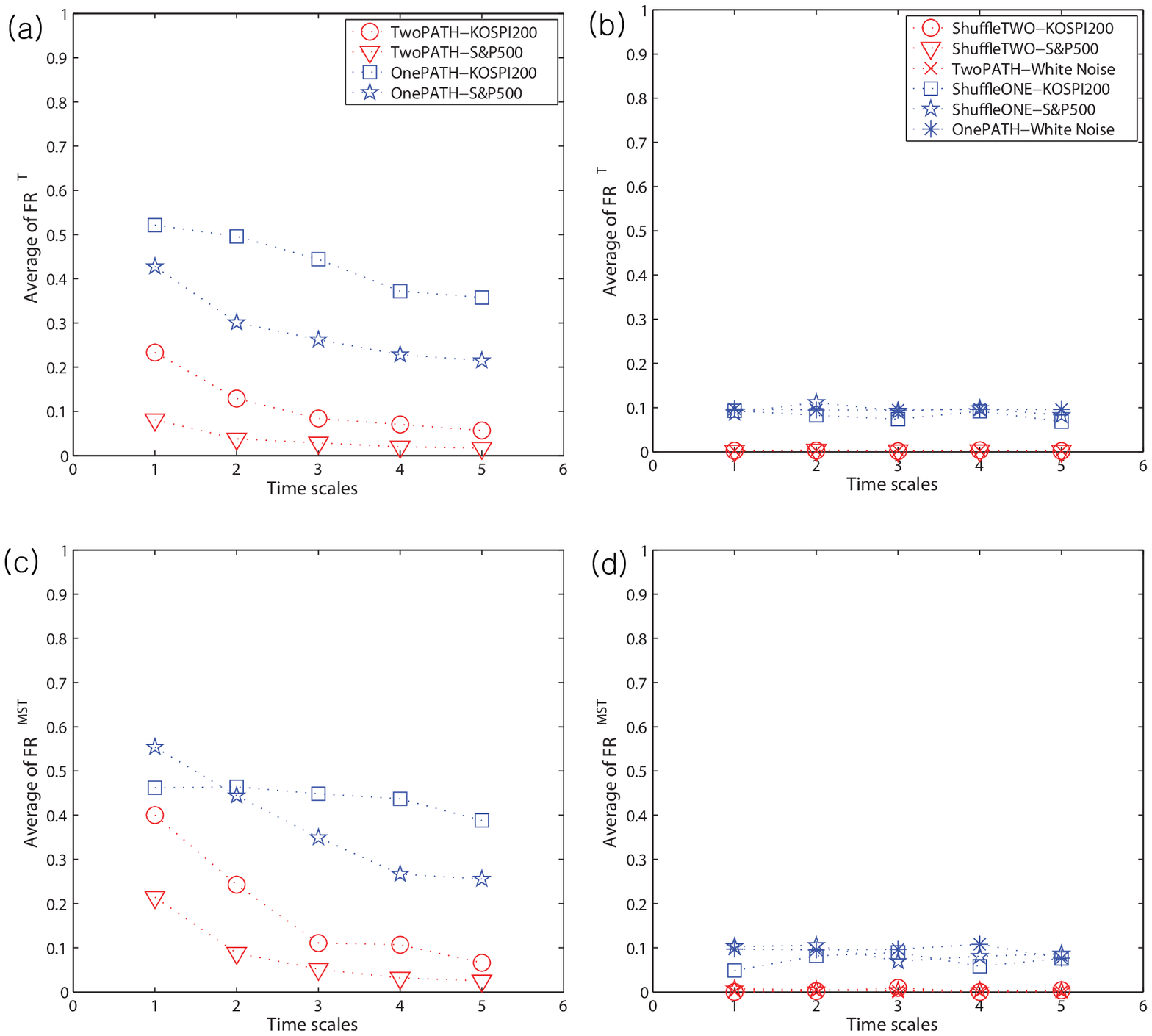}
\caption{(Online color). This figure shows the results of the effects on the information flow among stocks according to the changes in the time scale of return. Figs. 1(a) and (b) show the average ratio of significant information flow among stocks, observed using all $N(N-1)/2$ links among stocks, otherwise, Figs. 1(c) and (d) observed using $N-1$ links reduced via the MST method. Significant mutual exchange of information flow is designated by red circles for the Korean stock market and red triangles for the US stock market, and significant one-way direction of information flow with blue squares (Korea) and blue pentagrams (US). Meanwhile, the results obtained using shuffled time series and white noise data are provided in Figs. 1 (b) and (d).}
\label{fig:1}
\end{figure}

\newpage
\clearpage

\begin{figure}
\includegraphics[width=1.0\textwidth]{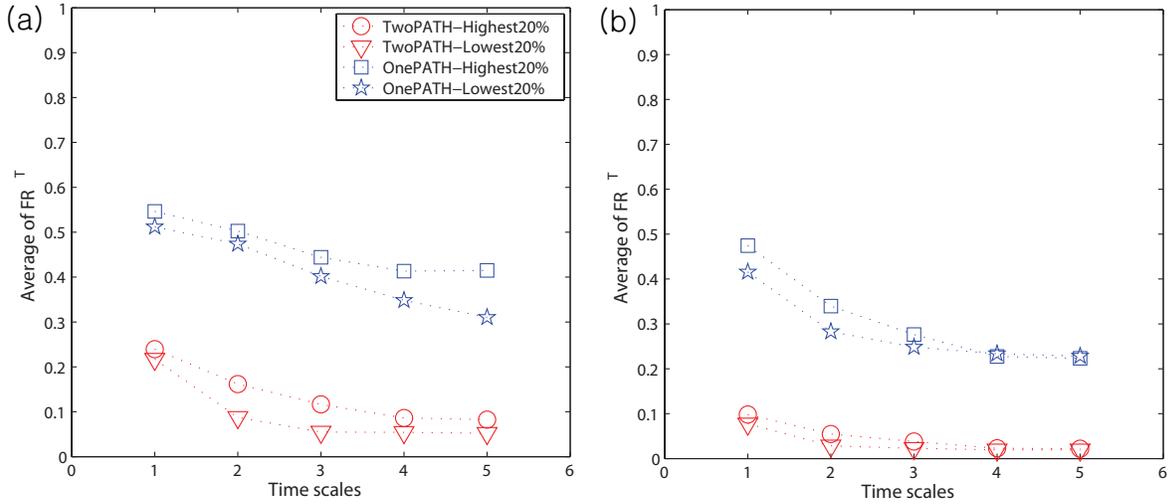}
\caption{(Online color). This figure displays results of the comparison between the amounts of significant information flow between the group with the upper 20\% correlation and that with the lower 20\% correlation. Figs. 2(a) and (b) show the results using data from the Korean and US stock markets, respectively. In the group with the upper 20\% correlation, significant mutual exchange and one-way direction of information flow are designated with red circles and blue squares, respectively, and the group with the lower 20\% correlation is indicated by red triangles and blue pentagrams. }
\label{fig:2}
\end{figure}

\newpage
\clearpage

\begin{figure}
\includegraphics[width=1.0\textwidth]{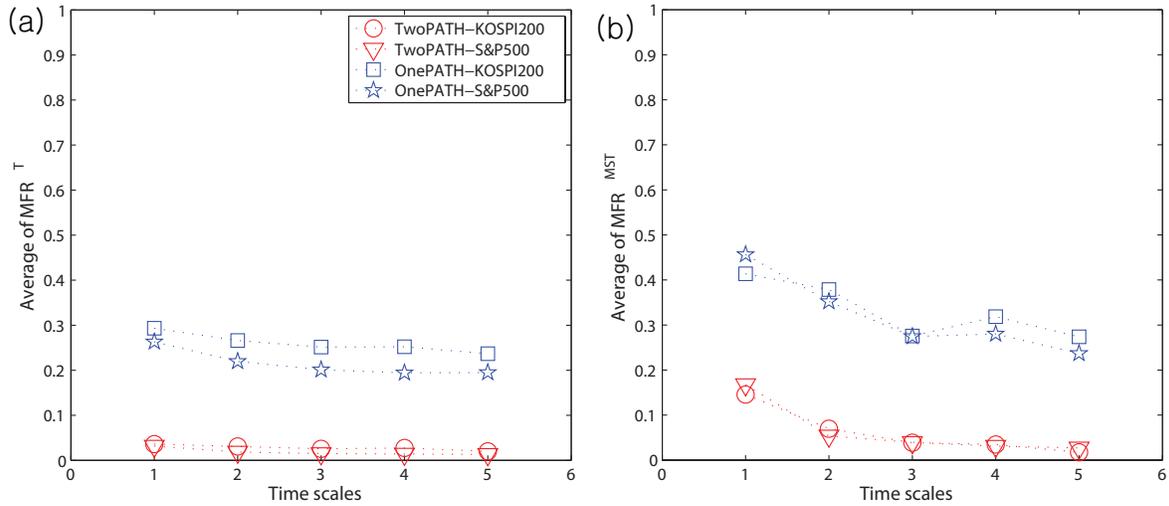}
\caption{(Online color). This figure displays the results of information flow among stocks observed using the return data without the market factor properties. Figs. 3(a) and (b) are results using $N(N-1)/2$ links and $N-1$ links via the MST method, respectively. The figure shows the average ratios of significant information flows. The average ratios of mutual exchange of information flow are designated with red circles (Korean stock market data) and red triangles (US stock market data), and those of one-way direction of information flow with blue squares (Korea) and blue pentagrams (US).}
\label{fig:3}
\end{figure}

\newpage
\clearpage

\begin{figure}
\includegraphics[width=1.0\textwidth]{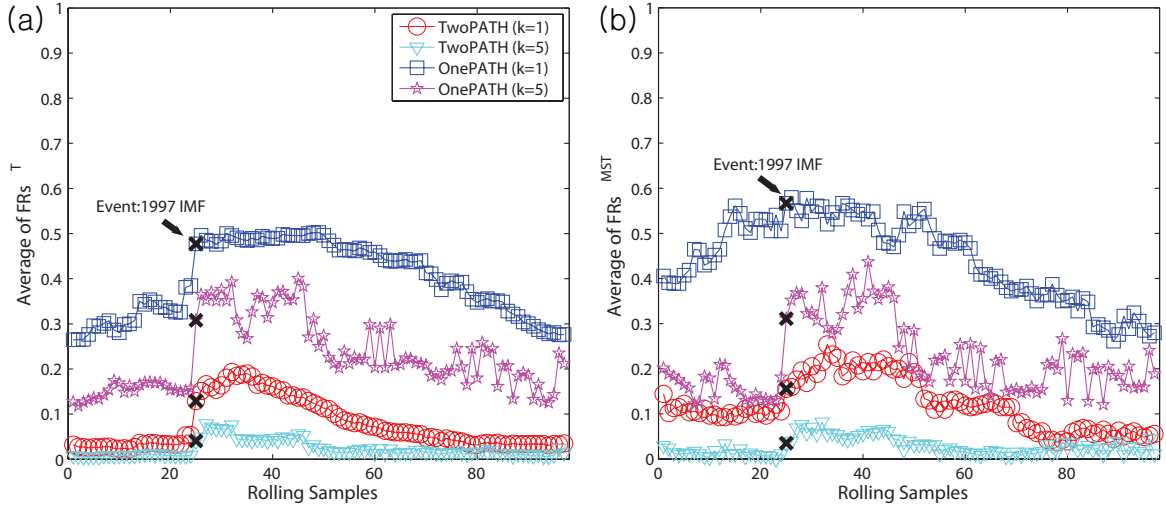}
\caption{(Online color). This figure shows the results of our examination of the effects on the information flow among stocks according to changes in market status. Figs. 4(a) and (b) show the results using $N(N-1)/2$ links and $N-1$ links via the MST method, respectively. The figure shows the average ratio of significant mutual exchange and one-way direction of information flow for the two time scales of return of $k=1$ and $k=5$. The mutual exchange and one-way direction of information flow for $k=1$ are designated with red circles and blue squares, respectively, and with cyan triangles and magenta pentagrams, respectively for $k=5$. The Asia Foreign Exchange Crisis of December 1997 is indicated by an `x'.}
\label{fig:4}
\end{figure}

\newpage
\clearpage

\begin{table}
\begin{center}
\begin{tabular}{|c|c|c|c|c|c|c|}
\hline
\multicolumn{2}{|c|}{}&\multicolumn{5}{|c|}{Time Scales} \\
\cline{3-7}
\multicolumn{2}{|c|}{}&$k=1$&$k=2$&$k=3$&$k=4$&$k=5$ \\
\hline
\multicolumn{7}{|l|}{\textbf{CASE 1}: using $N(N-1)/2$ links among stocks} \\
\hline
Korean&Avg. of $DFR(k,L:A\rightleftharpoons B)^T$&0.1969&0.0980&0.0580&0.0431&0.0366\\
\cline{2-7}
Stocks&Avg. of $DFR(k,L:A\rightarrow B)^T$&0.2280&0.2298&0.1923&0.1199&0.1207\\
\hline
US&Avg. of $DFR(k,L:A\rightleftharpoons B)^T$&0.0495&0.0194&0.0137&0.0057&0.0037\\
\cline{2-7}
Stocks&Avg. of $DFR(k,L:A\rightarrow B)^T$&0.1643&0.0809&0.0612&0.0338&0.0198\\
\hline
\multicolumn{7}{|l|}{\textbf{CASE 2}: using $N-1$ links among stocks} \\
\hline
Korean&Avg. of $DFR(k,L:A\rightleftharpoons B)^{MST}$&0.2544&0.1728&0.0719&0.0718&0.0486\\
\cline{2-7}
Stocks&Avg. of $DFR(k,L:A\rightarrow B)^{MST}$&0.0485&0.0855&0.1728&0.1184&0.1146\\
\hline
US&Avg. of $DFR(k,L:A\rightleftharpoons B)^{MST}$&0.0466&0.0343&0.0123&0.0007&-0.0020\\
\cline{2-7}
Stocks&Avg. of $DFR(k,L:A\rightarrow B)^{MST}$&0.0977&0.0919&0.0750&-0.0136&0.0188\\
\hline

\end{tabular}
\caption{\label{table:1}This table displays the differences of average ratios $\overline{DFR(.)}=\overline{FR(.)}-\overline{MFR(.)}$ of mutual exchange and one-way direction of information flow observed in Fig. 1 and Fig. 3, respectively, according to $N(N-1)/2$ and $N-1$ links as well as the time scale $k$.}
\end{center}
\end{table}

\end{document}